\documentclass[fleqn,usenatbib,useAMS]{mnras}
\usepackage{natbib}
\usepackage{epsf}
\usepackage{color}

\usepackage{graphicx}	
\usepackage{amsmath}	
\usepackage{multicol}        
\usepackage{bm}		
\usepackage{pdflscape}	
\usepackage{auto-pst-pdf}

\newcommand{\Msun}{\ifmmode\mbox{M}_{\odot}\else$\mbox{M}_{\odot}$\fi}
\newcommand{\Rsun}{\ifmmode\mbox{R}_{\odot}\else$\mbox{R}_{\odot}$\fi}
\newcommand{\Mearth}{\ifmmode\mbox{M}_{\oplus}\else$\mbox{M}_{\oplus}$\fi}
\newcommand{\Rearth}{\ifmmode\mbox{R}_{\oplus}\else$\mbox{R}_{\oplus}$\fi}
\newcommand{\Mp}{M_{\rm p}}
\newcommand{\Mc}{M_{\rm c}}

\newcommand{\msG}{\mathcal{G}}
\newcommand{\msF}{\mathcal{F}}
\newcommand{\msV}{\mathcal{V}}

\def\be{\begin{equation}}
\def\ee{\end{equation}}
\newcommand{\bb}{\begin{bmatrix}}
\newcommand{\eb}{\end{bmatrix}}

\usepackage{deluxetable}

\usepackage[T1]{fontenc}
\usepackage{ae,aecompl}

\usepackage{newtxtext,newtxmath}
 
\title{Tests of Gravitational Symmetries with Pulsar Binary J1713+0747}

\author[W.~W.~Zhu et~al.]{\parbox{\textwidth}{
W.~W.~Zhu,$^{1,2}$\thanks{E-mail:zhuww@mpifr-bonn.mpg.de}
G.~Desvignes,$^{1}$
N.~Wex,$^{1}$
R.~N.~Caballero,$^{1}$
D.~J.~Champion,$^{1}$
P.~B.~Demorest,$^{3}$
J.~A.~Ellis,$^{4}$
G.~H.~Janssen,$^{5,6}$
M.~Kramer,$^{1,7}$
A.~Krieger,$^{1}$
L.~Lentati,$^{8}$
D.~J.~Nice,$^{9}$
S.~M.~Ransom,$^{10}$
I.~H.~Stairs,$^{11}$
B.~W.~Stappers,$^{7}$
J.~P.~W.~Verbiest,$^{1,12}$
Z.~Arzoumanian,$^{13,14}$
C.~G.~Bassa,$^{5}$
M.~Burgay,$^{15}$
I.~Cognard,$^{16,17}$
K.~Crowter,$^{11}$
T.~Dolch,$^{18}$
R.~D.~Ferdman,$^{19}$
E.~Fonseca,$^{20}$
M.~E.~Gonzalez,$^{11,21}$
E.~Graikou,$^{1}$
L.~Guillemot,$^{16,17}$
J.~W.~T.~Hessels,$^{22,5}$
A.~Jessner,$^{1}$
G.~Jones,$^{23}$
M.~L.~Jones,$^{24,31}$
C.~Jordan,$^{7}$
R.~Karuppusamy,$^{1}$
M.~T.~Lam,$^{24,31}$
K.~Lazaridis,$^{1}$
P.~Lazarus,$^{1}$
K.~J.~Lee,$^{25,1}$
L.~Levin,$^{7}$
K.~Liu,$^{1}$
A.~G.~Lyne,$^{7}$
J.~W.~McKee,$^{1,7}$
M.~A.~McLaughlin,$^{24,31}$
S.~Os{\l}owski,$^{1,12,30}$
T.~Pennucci,$^{24,26}$
D.~Perrodin,$^{15}$
A.~Possenti,$^{15}$
S.~Sanidas,$^{22,7}$
G.~Shaifullah,$^{1,12,5}$
R.~Smits,$^{5}$
K.~Stovall,$^{27}$
J.~Swiggum,$^{28}$
G.~Theureau,$^{16,17,29}$
C.~Tiburzi,$^{1,12}$  
}
\vspace{0.4cm} \\
\parbox{\textwidth}{
$^{1}$ Max-Planck-Institut f\"ur Radioastronomie, Auf dem H\"ugel 69, D-53121 Bonn, Germany\\
$^{2}$ National Astronomical Observatories, Chinese Academy of Science, 20A Datun Road, Chaoyang District, Beijing 100012, China\\
$^{3}$ National Radio Astronomy Observatory, P.~O.~Box 0, Socorro, NM, 87801, USA\\
$^{4}$ Jet Propulsion Laboratory, California Institute of Technology, 4800 Oak Grove Dr. Pasadena CA, 91109, USA\\
$^{5}$ ASTRON, the Netherlands Institute for Radio Astronomy, Postbus 2, 7990 AA, Dwingeloo, The Netherlands\\
$^{6}$ Department of Astrophysics/IMAPP, Radboud University, P.O. Box 9010, 6500 GL Nijmegen, The Netherlands\\
$^{7}$ Jodrell Bank Centre for Astrophysics, School of Physics and Astronomy, The University of Manchester, Manchester M13 9PL, UK\\
$^{8}$ Institute of Astronomy / Battcock Centre for Astrophysics, University of
Cambridge, Madingley Road, Cambridge CB3 0HA, United Kingdom\\
$^{9}$ Department of Physics, Lafayette College, Easton, PA 18042, USA\\
$^{10}$ National Radio Astronomy Observatory, Charlottesville, VA 22903, USA\\
$^{11}$ Department of Physics and Astronomy, 6224 Agricultural Road, University of British Columbia, Vancouver, BC, V6T 1Z1, Canada\\
$^{12}$ Fakult\"at f\"ur Physik, Universit\"at Bielefeld, Postfach 100131, 33501 Bielefeld, Germany\\
$^{13}$ Center for Research and Exploration in Space Science and Technology and X-Ray Astrophysics Laboratory, NASA Goddard Space Flight Center, Code 662, Greenbelt, MD 20771, USA\\
$^{14}$ Universities Space Research Association, Columbia, MD 21046, USA\\
$^{15}$ INAF - ORA - Osservatorio Astronomico di Cagliari, via della Scienza 5, I-09047 Selargius (CA), Italy\\
$^{16}$ Laboratoire de Physique et Chimie de l'Environnement et de l'Espace LPC2E CNRS-Universit\'{e} d'Orl\'{e}ans, F-45071 Orl\'{e}ans, France\\
$^{17}$ Station de radioastronomie de Nan\c{c}ay, Observatoire de Paris, CNRS/INSU F-18330 Nan\c{c}ay, France\\
$^{18}$ Department of Physics, Hillsdale College, 33 E. College Street, Hillsdale, Michigan 49242, USA\\
$^{19}$ Faculty of Science, University of East Anglia, Norwich Research Park, Norwich NR4 7TJ, UK\\
$^{20}$ Department of Physics, McGill University, Montreal, QC H3A 2T8, Canada\\
$^{21}$ Department of Nuclear Medicine, Vancouver Coastal Health Authority, Vancouver, BC V5Z 1M9, Canada\\
$^{22}$ Anton Pannekoek Institute for Astronomy, University of Amsterdam, Science Park 904, 1098 XH Amsterdam, The Netherlands\\
$^{23}$ Department of Physics, Columbia University, 550 W. 120th St. New York, NY 10027, USA\\
$^{24}$ Department of Physics and Astronomy, West Virginia University, P.O. Box 6315, Morgantown, WV 26506, USA\\
$^{25}$ Kavli institute for astronomy and astrophysics, Peking University, Beijing 100871, P.R. China\\
$^{26}$ Hungarian Academy of Sciences MTA-ELTE Extragalactic Astrophysics Research Group, Institute of Physics, E\"{o}tv\"{o}s Lor\'{a}nd University, P\'{a}zm\'a{a}ny P. s. 1/A, Budapest 1117, Hungary.\\
$^{27}$ Department of Physics and Astronomy, University of New Mexico, Albuquerque, NM, 87131, USA\\
$^{28}$ Center for Gravitation, Cosmology and Astrophysics, Department of Physics, University of Wisconsin-Milwaukee, P.O. Box 413, Milwaukee, WI 53201, USA\\
$^{29}$ LUTH, Observatoire de Paris, PSL Research University, CNRS, Universit\'e Paris Diderot, Sorbonne Paris Cit\'e, F-92195 Meudon, France\\
$^{30}$ Centre for Astrophysics and Supercomputing, Swinburne University of Technology, Hawthorn Vic 3122, Australia\\
$^{31}$ Center for Gravitational Waves and Cosmology, West Virginia University, Chestnut Ridge Research Building, Morgantown, WV 26505, USA\\
}}

\begin{document}
\label{firstpage}
\pagerange{\pageref{firstpage}--\pageref{lastpage}}

\maketitle

\clearpage
 
\begin{abstract}
Symmetries play an important role in modern theories of gravity. The strong
equivalence principle (SEP) constitutes a collection of gravitational symmetries
which are all implemented by general relativity. Alternative theories, however, 
are generally expected to violate some aspects of SEP. We test three aspects of 
SEP using observed change rates in the orbital period and eccentricity of binary 
pulsar J1713+0747: 1.\ the gravitational constant's constancy as part of locational 
invariance of gravitation; 2.\ the post-Newtonian parameter $\hat{\alpha}_3$ in 
gravitational Lorentz invariance; 3.\ the universality of free fall (UFF) for 
strongly self-gravitating bodies. Based on the pulsar timing result of the combined 
dataset from the North American Nanohertz Gravitational Observatory (NANOGrav) and 
the European Pulsar Timing Array (EPTA), we find $\dot{G}/G = (-0.1 \pm 0.9) \times 
10^{-12}\,{\rm yr}^{-1}$, which is somewhat weaker than Solar system limits, but applies
for strongly self-gravitating objects. Furthermore, we obtain the constraints $|\Delta|< 
0.002$ for the UFF test and $-3\times10^{-20} < \hat{\alpha}_3 < 4\times10^{-20}$ at 95\% 
confidence. These are the first direct UFF and $\hat{\alpha}_3$ tests based on pulsar binaries, and they overcome various limitations of previous tests.
\end{abstract}

\begin{keywords}
pulsars: individual (PSR J1713+0747) --- Radio: stars --- stars: neutron
--- Binaries:general --- gravitation -- relativity
\end{keywords}


\section{Introduction}
\label{sec:intro}

Einstein's equivalence principle (EEP) is one of the guiding ideas that aided Einstein to 
conceive the theory of general relativity (GR). EEP states that non-gravitational 
experiments in a local Lorentz frame should give the same result regardless of when 
and where they take place. This principle helped in establishing the idea that 
gravity is the manifestation of curved spacetime, which can be abstracted as a 
four-dimensional manifold endowed with a Lorentzian metric, where freely falling test 
bodies follow geodesics of that metric (universality of free fall), and the local non-gravitational laws of physics are those of special relativity. Gravity theories built upon 
this concept are called ``metric theories of gravity'', like GR and scalar-tensor 
theories. See \citet{Will93} for details.

The strong equivalence principle (SEP) extends the EEP by including local 
gravitational aspects of the test system \citep{will14}. The universality of free fall is 
extended to self-gravitating bodies, which fall in an external gravitational field. 
Furthermore, local test experiments, including gravitational ones, should give the same 
results regardless of the location or velocity of the test system. It is conjectured 
that GR is the only gravity theory that fully embodies SEP.\footnote{There is 
one known exception, which however is falsified by Solar system experiments 
\citep{der11}} Although in metric theories of gravity all matter fields couple only to 
one physical metric (``universal coupling''), alternatives to GR generally introduce 
auxiliary gravitational fields (e.g.\ one or more scalar fields) which ultimately lead to 
a violation of SEP at some point. For this reason, testing the symmetries related to SEP 
has strong potential to either exclude (or tightly constrain) alternative gravity theories 
or falsify GR. It is therefore a powerful tool in searching for new physics. To date, all experimental evidence supports SEP and, therefore, GR \citep{will14,sw16}. 

The post-Newtonian parameterization (PPN) is a formalism introduced by \citet{tw71} and 
\citet{wn72} to describe generically the potential deviation from GR in metric 
theories of gravitation at the post-Newtonian level. Through a set of simple assumptions, 
such as slow-motion, weak field, and no characteristic length scales in the gravitational 
interaction, the PPN formalism can encompass most metric theories using only ten 
parameters. Most of these PPN parameters (or combinations of them) are directly 
related to a violation of specific aspects of SEP. For strongly self-gravitating bodies, 
like neutron stars, these PPN parameters become kind of body dependent quantities, which 
are functions of the compactness of the bodies of the system \citep[see e.g.][]{dam09}.
Hence, one can have situations where a theory is in (nearly) perfect agreement with GR in 
the Solar system, but shows significant violations of SEP in the presence of strongly 
self-gravitating bodies. A particularly extreme example is spontaneous scalarization, which is 
a non-perturbative strong-gravity effect that is known for certain scalar-tensor theories 
\citep{de93}. 

Alternative theories of gravity, generally, also predict a temporal change in the locally 
measured Newtonian gravitational constant $G$, which is caused by the expansion of the Universe 
\citep{Will93,uzan11}. Such a change in the local gravitational constant constitutes a 
violation of the local position invariance, which also refers to position in time. Hence, a
varying $G$ directly violates one of the three main pillars of SEP. One of the testable 
consequences of a change in $G$ are changes in the orbital parameters of the Solar system and 
binary systems, in particular the size of an orbit and the orbital period. Again, the situation is more complicated in the presence of strongly self-gravitating bodies \citep{nor90,wex14}.

Some alternative theories of gravity violate SEP by introducing a preferred frame of reference for the gravitational interaction. Generally, this preferred frame can be identified with the global mass distribution in the universe, which is the frame in which the cosmic microwave background is isotropic, i.e.\ has no dipole. In the PPN formalism, there are three parameters, $\alpha_1$, $\alpha_2$, and $\alpha_3$, which are related to such kinds of symmetry breaking. The parameter $\alpha_3$ is linked to two effects, a preferred frame effect and a violation of conservation of total momentum \citep{Will93}. In this paper, we test the PPN parameter $\alpha_3$ through the fact that it causes an anomalous self-acceleration of a spinning body, which is proportional to and perpendicular to the object's spin and motion with respect to the preferred frame. Such acceleration would lead to an observable effect in a binary system, such as an anomalous drift in the eccentricity of the binary. PSR J1713+0747, thanks to its high spin frequency and measurable proper motion, has the best figure of merit for testing $\alpha_3$ in the present.
More precisely, with binary pulsars one tests the quantity $\hat{\alpha}_3$, which is a 
generalization of $\alpha_3$ to a situation with strongly self-gravitating objects. Therefore, $\hat{\alpha}_3$ also contains preferred frame effects related to the binding energy of the neutron star \citep[cf.\ discussion in][]{de92a,will18}

Some pulsar binary systems are particularly useful for certain tests of SEP \citep[see][for a 
recent review]{sw16}. In this paper, we measure the change rate in the orbital periodicity and 
eccentricity of the pulsar-white dwarf binary PSR~J1713+0747 and use that to test the following 
three aspects of SEP: 1. the gravitational constant's constancy; 2. the PPN parameter 
$\hat{\alpha}_3$ in the context of Lorentz invariance of gravitation and conservation of 
total momentum; 3. the universality of free fall (UFF) for strongly self-gravitating bodies.
Section \ref{sec:timing} describes the pulsar timing, Section \ref{sec:gdot}, \ref{sec:sep}, 
\ref{sec:alpha3} describe the test for the gravitational constant's constancy, the universality 
of free fall test, and the test for $\hat{\alpha}_3$. In Section \ref{sec:summary} we give the 
conclusion and the summary.


\section{Method and Results}
\label{sec:method}

PSR J1713+0747 is a millisecond pulsar orbiting a 0.29\,$\Msun$ white
dwarf. The pulsar's short spin period (4.5~ms) and narrow pulse profile enable
us to measure its pulse time of arrivals (TOAs) at sub-microsecond precision. This pulsar is monitored by both the North American Nanohertz Gravitational
Observatory (NANOGrav), the European Pulsar Timing Array (EPTA) and the Parkes Pulsar Timing Array (PPTA) for the purpose of detecting nHz gravitational waves. 
A major part of the data used in this work come from the NANOGrav\footnote{\url{www.nanograv.org}} and EPTA programs. 
The EPTA\footnote{\url{www.epta.eu.org}} is a collaboration of European
institutes to work towards the direct detection of low-frequency gravitational waves and the running of the Large European Array for Pulsars (LEAP).
These data were taken between 1993 and 2014 using observatories 
including William~E.~Gordon Telescope of Arecibo observatory, Robert~C.~Byrd Green Bank Telescope, Effelsberg Telescope, Lovell Telescope of Jodrell Bank observatory, Nan\c{c}ay Radio Telescope, and Westerbork Synthesis Radio Telescope.
\citet{sns+05} published the early dataset for J1713+0747 from Arecibo and Green Bank Telescope (GBT). Subsequently, \citet{zsd+15} published the combined J1713+0747 data from \cite{sns+05} and NANOGrav 9yr data release \citep{abb+15}. \citet{dcl+16} published the data and timing analysis for the EPTA pulsars including J1713+0747.   
This work presents the first timing analysis of the combined J1713+0747 data published in \cite{zsd+15} and \cite{dcl+16}. 
We are able to model very accurately the binary system's orbit through various time-delaying effects like R\"{o}mer delay, Shapiro delay, and annual-orbital parallax. From these we can measure the masses of the
two stars, the sky orientation and inclination angle of the orbit, and the
position and proper motion of the system. The modeling was performed using the
pulsar timing software \textsc{Tempo2} \citep{ehm06} and the best-fit parameters are listed in Table~\ref{tab:pars}. 


\subsection{Pulsar Binary Timing}
\label{sec:timing}

For the timing of PSR~J1713+0747, we employ an extended version of the {\it 
ELL1}\footnote{The name ELL1 comes from the fact that eccentricity ($e$) is much less (LL: 
less less) than one (1).} pulsar binary model \citep{lcw+01}, which is valid for 
$e \equiv |\mathbf{e}| \ll 1$, where
$\mathbf{e}$ is the eccentricity vector of the orbit. {\it ELL1} models a pulsar 
binary orbit with small eccentricity by decomposing $\mathbf{e}$ into two orthogonal 
vectors $e_x$ and $e_y$, where $e_x \equiv e\cos\omega$ and $e_y \equiv e\sin\omega$,  
and $\omega$ is the longitude of periastron, i.e.\ the angle between $\mathbf{e}$ and the 
ascending node. We use $e_x$ to represent the component of $\mathbf{e}$ pointing from the centre of the orbit 
to the ascending node and $e_y$ represents the part pointing from Earth to the pulsar. 
\citet{lcw+01} showed that the R\"{o}mer delay of a small eccentricity orbit can be 
expressed simply as $\Delta_{R} = x[\sin\phi + (e_y/2)\sin2\phi - (e_x/2)\cos2\phi]$, omitting
higher order terms proportional to ${\cal O} (xe^2)$. Here $x$ is the
projected semi-major axis of the pulsar orbit in units of light-seconds, and $\phi \equiv 
n_{\rm b}(T - T_{\rm asc})$, where $n_{\rm b} = 2\pi/P_{\rm b}$ is the orbital frequency, 
$T$ the time of the pulsar and $T_{\rm asc}$ the so-called time of the ascending node (see 
\cite{lcw+01} for details). However, we find that the precision of this expression is 
insufficient for modeling PSR~J1713+0747's R\"{o}mer delay due to its timing precision. 
To increase the precision of our timing model, we extend the {\it ELL1} model by including the second order terms:
\begin{align}
  \Delta_R =
    & x \Big(\sin\phi-\frac{e_x}{2}\cos2\phi+\frac{e_y}{2}\sin2\phi\Big) 
    \nonumber\\
    & -\frac{x}{8}\Big( 5e_x^2\sin\phi - 3e_x^2\sin3\phi -2e_xe_y \cos\phi 
    \nonumber\\
    & + 6e_xe_y\cos3\phi + 3e_y^2\sin\phi + 3e_y^2\sin3\phi \Big)
      + {\cal O}(xe^3) \,.
  \label{eq:romer}
\end{align}
This extended {\it ELL1} model ({\it ELL1+}) is sufficient for modeling PSR~J1713+0747 since its 
$xe^3 \sim 0.01$\,ns. Furthermore, we express $\mathbf{e}$ as a function of time
[$e_x(t) = e_x(t_0) + \dot{e}_x t$ and $e_y(t) = e_y(t_0) + \dot{e}_y t$] to model the 
effect of a changing eccentricity, where $\dot{e}_x$ and $\dot{e}_y$ represent the change rate of $\mathbf{e}$ in time. The higher order terms of eq.~(\ref{eq:romer}) could then 
straightforwardly be added to the existing implementation of the {\it ELL1} model in 
TEMPO2 \citep{ehm06}. 

Apart from the changes in the binary model, the rest of the timing analysis 
follows those in \cite{zsd+15} and \cite{dcl+16}. For a high-timing precision pulsar such 
as PSR~J1713+0747, it is necessary to employ a comprehensive noise model including 
dispersion measure (DM) variation, jitter noise, and red noise. Here we use two 
different approaches. The first based on the noise analysis technique described in \cite{vl13, ell13, dlc+14, zsd+15, abb+15} and \cite{dec+16}. In \cite{dec+16}, the PSR J1713+0747 noise model enabled the strongest PTA-based $\nu$Hz gravitational limit. Our analysis uses the {\it DMX} model that groups 
TOAs into epochs and fit for DM for each group respectively. We model jitter noise as a correlated noises 
between TOAs from the same observations and the red noise as a stationary Gaussian process 
with a power law spectrum. This analysis is conducted using the \textsc{PAL2} software 
package\footnote{\url{https://github.com/jellis18/PAL2}} \citep{PAL2}. The second
approach is similar to the first one excepts that it models DM
variation as a power-law Gaussian process \citep{lah+13}. This analysis is 
conducted using the \textsc{TempoNest}\footnote{\url{https://github.com/LindleyLentati/TempoNest}} 
software package \citep{lah+14}. We find consistent results in the best-fit timing and noise model from both approaches. 
In this paper, we are mostly interested in testing theories of gravitation. Therefore, we choose to base our GR tests on the result from the first approach (presented in Table \ref{tab:pars}) because it yields slightly more conservative uncertainty on $\dot{\mathbf e}$. 

We use the solar system ephemeris DE421 \citep{fwb09} in the timing analysis instead of the more recent DE436 \citep{fp16}. \citet{abb+18a} showed that using DE436 leads to some marginally different timing results from using DE421. The discrepancies in different solar system ephemerides are in the masses and orbits of the outer solar system bodies, they cause extra timing residuals in time scales of the orbital periods of these bodies. PSR J1713+0747's orbital period is substantially smaller than those solar system bodies. Therefore, we argue that using different solar system ephemerides will have only marginal impact on our primary parameters of interest: $\dot{P}_{\rm b}$ and $\dot{\mathbf{e}}$.    


\begin{table}
\caption{ Timing model parameters\tablenotemark{a}\,~ from \textsc{Tempo2}. }
\label{tab:pars}
\begin{tabular}{lc}
\hline
\tabletypesize{\scriptsize}
Parameter & Best-fit values\\
\hline
\textit{Measured Parameters} & \\[1mm]
Right Ascension, $\alpha$ (J2000) & 17:13:49.5320247(9) \\
Declination,     $\delta$ (J2000) &  7:47:37.50612(2) \\
Proper motion in $\alpha$, $\mu_\alpha=\dot\alpha\cos\delta$ (mas\,yr$^{-1}$) & 4.918(3)\\
Proper motion in $\delta$, $\mu_\delta=\dot\delta$ (mas\,yr$^{-1}$)& $-$3.915(5)\\
Parallax, $\uppi$ (mas) & 0.87(4)\\
Spin Frequency, $\nu$\,(s$^{-1}$)      & 218.8118438547250(3)\\
Spin down rate, $\dot{\nu}$ (s$^{-2}$) & $-4.08379(4) \times 10^{-16}$\\
Dispersion Measure\tablenotemark{b}~ (pc\,cm$^{-3}$)&  15.970\\
Orbital Period, $P_{\rm b}$ (day)&  67.8251299228(5)\tablenotemark{c} \\
Change rate of $P_{\rm b}$, $\dot{P}_{\rm b}$ ($10^{-12}\,{\rm s\,s^{-1}}$)&  0.34(15)\\
$\hat{x}$ component of the eccentricity, $e_x$ & $-$0.0000747752(7) \\
$\hat{y}$ component of the eccentricity, $e_y$ &    0.0000049721(19) \\
Change rate of $e_x$, $\dot{e}_x$ (${\rm s}^{-1}$) & $ 0.4(4) \times 10^{-17}$ \\
Change rate of $e_y$, $\dot{e}_y$ (${\rm s}^{-1}$) & $-1.7(4) \times 10^{-17}$ \\
Time of ascending node, $T_{\rm asc}$ (MJD) & 53727.836759558(6) \\
Projected semi-major axis, $x$ (lt-s) & 32.34242184(12) \\
Orbital inclination, $i$ (deg) & 71.69(19) \\
Companion Mass, $M_c/\Msun$ & 0.290(11) \\
Position angle of ascending node, $\Omega$ (deg) & 89.7(6) \\
Profile frequency dependency parameter, FD1\tablenotemark{d} &  $-$0.00016376(18) \\
Profile frequency dependency parameter, FD2\tablenotemark{d} &     0.0001363(3) \\
Profile frequency dependency parameter, FD3\tablenotemark{d} &  $-$0.0000672(6) \\
Profile frequency dependency parameter, FD4\tablenotemark{d} &     0.0000152(5) \\
\hline
\textit{Fixed Parameters} & \\[1mm]
Solar System ephemeris & DE421 \\
Reference epoch for $\alpha$, $\delta$, and $\nu$ (MJD)&  53729 \\
Red Noise Amplitude      & $-$13.451 \\
Red Noise Spectral Index & $-$1.867 \\
\hline
\textit{Derived Parameters} & \\[1mm]
Intrinsic period derivative, $\dot{P}_{\rm Int}$(s~s$^{-1}$) & $8.96(3)\times10^{-21}$ \\
Pulsar mass, $\Mp/\Msun$ &  1.33(10) \\
Dipole magnetic field, $B$ (G)    & $2.048(3) \times 10^8$ \\
Characteristic age, $\tau_c$ (yr) & $8.08(3) \times 10^9$  \\
\hline
\tablenotetext{a}{We extend \textsc{Tempo2}'s {\it T2} binary model to include
higher order corrections from the {\it ELL1} model. Numbers in parentheses indicate 
the uncertainties on the last digit(s).  Uncertainties on parameters are estimated from 
the result of MCMC process in which the timing and noise model were evaluated.}
\tablenotetext{b}{The averaged DM value based on the DMX model. 
}
\tablenotetext{c}{Most pulsar timing model parameters presented in this paper are consistent with those reported in \citet{zsd+15} except for the orbital period $P_{\rm b}$. This is because $P_{\rm b}$ is defined differently in {\it ELL1} model used here from the {\it DD} model used in \citet{zsd+15}. In {\it DD} model, $P_{\rm b}$ is defined as the time between two periastron passings, while in {\it ELL1} model, $P_{\rm b}$ is the time between two ascending node passings.}
\tablenotetext{d} {See \citet{zsd+15} and \citet{abb+15} for the description and discussion of the FD model.}

\end{tabular}
\end{table}


\subsection{Testing the Time Variation of $G$}
\label{sec:gdot}

Through pulsar timing, we measure the apparent change rate of the binary's orbital period 
($\dot{P}_{\rm b}=(0.34\pm0.15)\times10^{-12}\,{\rm s\,s^{-1}}$, Table \ref{tab:pars}).
Despite being not statistically significant, the observed $\dot{P}_{\rm b}$ is consistent 
with what one expects from the apparent orbital period change caused by the binary's 
transverse motion (a.k.a.\ Shklovskii effect; \citealt{shk70}) and line-of-sight acceleration:
\begin{align}\label{eq:gdot}
  & \dot{P}_{\rm b}^{\rm Shk} = (\mu_{\alpha}^2+\mu_{\delta}^2)\frac{d}{c}P_{\rm b} 
                              = (0.65 \pm 0.03) \times 10^{-12}\,{\rm s\,s^{-1}}\,, \\
  & \dot{P}_{\rm b}^{\rm Gal} = \frac{A_{\rm G}}{c} P_{\rm b} 
                              = (-0.34 \pm 0.02) \times 10^{-12}\,{\rm s\,s^{-1}}\,.
\end{align}
Here, $\mu_{\alpha}$ and $\mu_{\delta}$ are the proper motion in right ascension and 
declination respectively, $d$ is the distance from timing parallax, $c$ is the speed of light, and $A_{G}$ is 
the system's line-of-sight acceleration computed using the Galactic potential in
\cite{McMillan16} (see discussion in Appendix~\ref{sec:appA}). Subtracting these two 
(external) contributions from the observed $\dot{P}_{\rm b}^{\rm Obs}$ yields the residual change rate of the orbital period
\begin{equation}
  \dot{P}_{\rm b}^{\rm Res} 
    = \dot{P}_{\rm b}^{\rm Obs} - \dot{P}_{\rm b}^{\rm Shk} - \dot{P}_{\rm b}^{\rm Gal} 
    = (0.03 \pm 0.15) \times 10^{-12}\,{\rm s\,s^{-1}}\,,
\end{equation}
which is consistent with the much smaller and undetectable intrinsic change 
$\dot{P}_{\rm b}^{\rm GR} = -6 \times 10^{-18}\,{\rm s\,s^{-1}}$ from quadrupolar 
gravitational radiation as predicted by GR.

This apparent consistency allows us to test the change rate of the (local) gravitational 
constant ($\dot{G}$) over the time span of the observation, since a $\dot{G}$ could lead 
to an observable change in $P_{\rm b}$ \citep{dgt88}, which has already been used to 
constrain $\dot{G}$ with binary pulsars \citep{dgt88, ktr94, nss+05, vbv+08, dvtb08, 
lwj+09, fwe+12, zsd+15}.

\cite{nor90} pointed out that a change of the gravitational constant also leads to 
changes in the neutron star's compactness and mass, and consequently leads to an additional 
contribution to $\dot{P}_{\rm b}$ which needs to be incorporated in our $\dot{G}$ tests with 
binary pulsars. If the companion is a weakly self-gravitating 
body, like in the case of PSR~J1713+0747, then one finds to leading order \citep[cf.\ eq.\ (18) in]{nor90}
\begin{equation}\label{eq:PbdotGdot}
  \dot{P}_{\rm b}^{\dot{G}} \simeq -2\frac{\dot{G}}{G} \left[
    1 - \frac{2\Mp + 3\Mc}{2(\Mp+\Mc)} s_{\rm p} 
      - \frac{2\Mc + 3\Mp}{2(\Mp+\Mc)} s_{\rm c} \right]P_{\rm b}\,,
\end{equation}
where $\Mp$, $\Mc$ are the pulsar mass and the companion mass, respectively.
The quantity $s_{\rm p}$ denotes the ``sensitivity'' of the neutron star 
and the white dwarf and are given by \citep[cf.][]{Will93}
\begin{equation}\label{eq:sp}
  s_{\rm p} \equiv -\left.\frac{\partial\ln \Mp}{\partial\ln G}\right|_N 
  \quad \mbox{and} \quad 
  s_{\rm c} \equiv -\left.\frac{\partial\ln \Mc}{\partial\ln G}\right|_N \,,
\end{equation}
respectively, where the number of baryons $N$ is held fixed. The sensitivity $s_{\rm p}$ 
of a neutron star depends on its mass, its equation of state (EoS), and the theory of gravity 
under consideration. As a reference, for Jordan-Fierz-Brans-Dicke (JFBD) gravity and the 
EoS AP4 in \cite{lp01} one finds for a $1.33\,\Msun$ neutron star, like 
PSR~J1713+0747, $s_{\rm p} \simeq 0.16$. Following \cite{de92}, we further assume, as a 
first order approximation, that $s_{\rm p}$ is proportional to the mass
\begin{equation} \label{eq:spMp}
  s_{ \rm p} = 0.16 \left( \frac{\Mp}{1.33\,\Msun} \right) \,.
\end{equation}
We will use this (simplified) relation in our generic calculations below but will keep in 
mind that depending on the EoS and the theory of gravity eq.~(\ref{eq:spMp}) might only be a rough 
estimate. Furthermore, it is important to 
note, that the usage of the sensitivity~(\ref{eq:sp}) and eq.~(\ref{eq:spMp}) comes with 
certain assumptions about how gravity can deviate from GR in the strong field of neutron 
stars. It is evident that such a description cannot capture non-perturbative strong-field 
effects, like those discussed by \cite{de93}. For a weakly self-gravitating body $A$, one has 
$s_A \simeq - E_A^{\rm grav}/M_Ac^2$, where $E_A^{\rm grav}$ is the gravitational binding 
energy of the body. Hence, one has $s_{\rm c} \simeq 3 \times 10^{-5}$ for the white-dwarf 
companion to PSR~J1713+0747 --- negligible in eq.~(\ref{eq:PbdotGdot}).

A time-varying gravitational constant generally indicates a violation of SEP. On 
the other hand, most gravitational theories that violate SEP also predict the 
existence of dipolar gravitational radiation (DGR). Such waves are very efficient in 
draining orbital energy from an (asymmetric) binary. The gravitational-wave 
damping due to DGR enters the equations-of-motion of a binary already at the 1.5 
post-Newtonian ($v^3/c^3$) level \citep[see e.g.][]{mw13}, and to leading order 
adds the following change to the orbital period
\begin{equation}\label{eq:PbdotD}
  \dot{P}_{\rm b}^{\rm D} \simeq 
    -\frac{2 G}{c^3}\,n_{\rm b}\,\frac{\Mp\Mc}{\Mp + \Mc}\,
     \kappa_{\rm D} (s_{\rm p} - s_{\rm c})^2 + {\cal O}(s_{\rm p}^3) \,,
\end{equation}
where $n_{\rm b} \equiv 2\pi/P_{\rm b}$ \citep{Will93}. The quantity $\kappa_{\rm D}$ is a 
body-independent constant, which depends on the fundamental parameters of the gravity 
theory under consideration. In JFBD gravity, for instance, one finds that
\begin{equation} \label{eq:kappaD_JFBD}
   \kappa_{\rm D} = \frac{2}{\omega_{\rm BD} + 2} \,,
\end{equation}
where $\omega_{\rm BD}$ is the Brans-Dicke parameter \citep{Will93}.\footnote{In JFBD 
gravity the effective scalar coupling $\alpha_A$ and the sensitivity $s_A$ of a neutron 
star are related by $\alpha_A = \alpha_0 (1 - 2 s_A)$ \citep[cf.\ Chapter 8 in][]{de92}.} 
For completeness, we have kept $s_{\rm c}$ in eq.~(\ref{eq:PbdotD}) although, as mentioned 
above, it is negligible in our case.

\begin{figure}
  \includegraphics[width=\columnwidth]{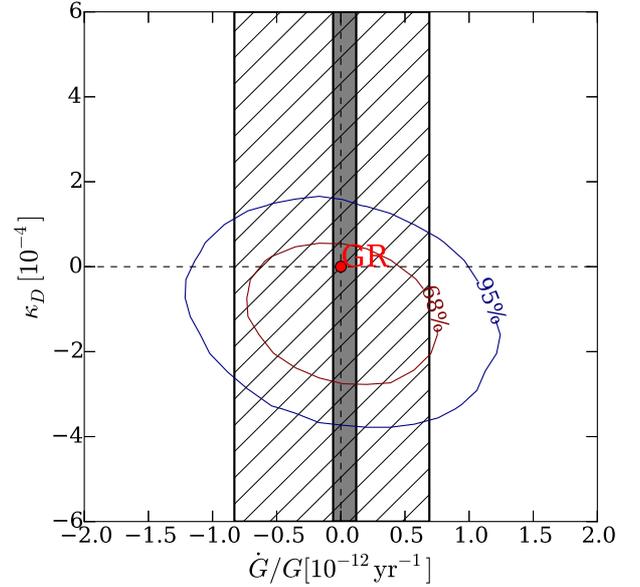} \\ 
  \caption{\label{fig:gdot} Confidence contours of $\dot{G}/G$ and $\kappa_{\rm D}$ 
     computed from MCMC simulations based on timing results of PSRs J0437$-$4715, 
     J1738+0333, and J1713+0747. The shaded area and gray area mark the 95\% confidence 
     limit from LLR \citep{hmb10} and planetary ephemerides \citep{fle+15}, respectively.} 
\end{figure}

As emphasized by \cite{lwj+09}, in a theory-agnostic approach the test of $\dot{G}$ and 
$\kappa_{\rm D}$ requires at least two pulsar binary systems with different 
orbital periods to break the degeneracy between the two contributions 
eq.~(\ref{eq:PbdotGdot}) and eq.~(\ref{eq:PbdotD}). 
This is because the extra variation in the orbital period due to dipolar gravitational radiation is stronger in binaries with shorter orbits ($\propto P_{\rm b}^{-1}$) while the that caused by $\dot{G}$ increases with orbital period ($\propto P_{\rm b}$). Therefore testing $\dot{G}/G$ using binaries of significantly different orbital periods breaks the degeneracy between the two effects \cite[see][for further details]{lwj+09}.
Here we adapt the method of \cite{lwj+09} and incorporate results from three different pulsar-white dwarf systems, namely PSRs J0437$-$4715 \citep{rhc+16}, and J1738+0333 \citep{fwe+12}, in combination with the results for PSR~J1713+0747 obtained in this work. Each pulsar provides a constraint on $\dot{P}_{\rm b}^{\rm D} + \dot{P}_{\rm b}^{\dot{G}}$, and hence via eqs.~(\ref{eq:PbdotGdot}) and (\ref{eq:PbdotD}) excludes certain regions in the $\dot{G}/G$--$\kappa_{\rm D}$ plane. As a result of the large difference in orbital period, these constraints are complementary and consequently lead to a small region of allowed values in the $\dot{G}/G$--$\kappa_{\rm D}$ plane (see Figure~\ref{fig:gdot}). The individual constraints on $\dot{G}$ and $\kappa_{\rm D}$ are
\begin{align}
  & \dot{G}/G = (-0.1 \pm 0.9) \times10^{-12}\,{\rm yr}^{-1} \,, \label{eq:GdotLim} \\
  & \kappa_{\rm D} = (-0.7 \pm 2.2) \times 10^{-4} \,.           \label{eq:kappaDLim}
\end{align}
They slightly improve the \cite{zsd+15} results based on NANOGrav-only 
PSR~J1713+0747 data.

\subsection{Testing the Universality of Free Fall for Strongly Self-gravitating Bodies}
\label{sec:sep}

One of the important pillars of SEP is the extension of the universality of free fall 
(UFF) to objects with significant gravitational binding energy $E^{\rm grav}$, i.e.\ the 
weak equivalence principle (WEP) is valid for test particles as well as for 
self-gravitating bodies. Every metric theory of gravity, by definition, 
fulfills WEP for test particles. On the other hand, alternatives to GR are usually
expected to violate WEP in the interaction of self-gravitating bodies 
\citep{Will93,will14}. According to such theories, objects with different binding energy 
feel different accelerations in an external gravitational field ${\bf g}$. More 
specifically, a binary system composed of two stars with different compactness would undergo
a ``gravitational Stark effect'' that polarizes the binary orbit in a characteristic way. 
In the Earth-Moon system, this is called the Nordtvedt effect and has been tightly 
constrained by Lunar Laser Ranging \citep{nor68,mhb12,wtb2012}. 
Pulsar binary systems falling in the gravitational field of our Galaxy would (slowly) 
oscillate between a more and less eccentric configuration. \citet{ds91} showed that the 
observed eccentricity is a combination of an intrinsic eccentricity and a forced 
eccentricity: 
\begin{equation}
  \textbf{\textit{e}}_F = \frac{\Delta \cdot {\bf g}_{\bot}c^2}
                          {2\msF\msG(\Mp+\Mc)n_{\rm b}^2}.
\end{equation}
Here $\msF$ is a theory-dependent (and "sensitivity"-dependent) factor that accounts for 
potential deviations from GR in the rate of periastron advance $\dot{\omega}$. By definition,
$\msF = 1$ in GR and indeed, it is constrained to be close to $1$ by observations. For instance, from the 
Double Pulsar PSR J0737-3039A/B \citep{bdp+03,lbk+04} one can quite generically infer that $|\msF - 1| \lesssim 10^{-3}$ 
\citep{kw09}. Therefore, we can safely assume $\msF=1$ in our analysis, in particular 
since the mass of PSR~J1713+0747 is comparable to the masses in the Double Pulsar. 
$\msG\approx G$ is the effective gravitational constant in the interaction between the 
pulsar and the white dwarf. The vector ${\bf g}_{\bot}$ is the projection of the Galactic 
acceleration ${\bf g}$ onto the orbital plane, and $\Delta$ is the fractional difference 
in the accelerations between the pulsar and white dwarf, and therefore a dimensionless 
measure of the significance of the UFF violation. 

\citet{ds91} have put forward a method to constrain $\Delta$ from small-eccentricity 
binary pulsars with white dwarf companions, utilizing in probabilistic considerations the 
smallness of the observed eccentricities. This so-called ``Damour-Sch\"afer'' test has 
been extended to make use of an ensemble of suitable pulsar-white dwarf systems 
\citep{wex97, sfl+05} of which the precise orbital orientations and proper motions are unknown. The currently best limits from this method are 
$|\Delta| < 5.6\times10^{-3}$ \citep{sfl+05} and $|\Delta| < 4.6\times10^{-3}$ 
\citep{gsf+11}\footnote{As discussed 
in detail in \cite{wex14}, the limit by \citet{gsf+11} comes with a caveat, it is slightly optimistic because of the inclusion of a pulsar unsuitable for the test.}. 
The validity and effectiveness of the Damour-Sch\"afer test rely on some 
(strong-field and probabilistic) assumptions. It does not improve with timing precision 
and is not capable of actually detecting a violation of the UFF 
\citep[see discussions in][]{dam09,fkw12}. For this reason, it is desirable to have a 
direct test with a single binary pulsar. \citet{fkw12} already identified PSR~J1713+0747 
as a potential candidate for such a test. 
In \citealt{zsd+15}, PSR~J1713+0747 had been used in a (single system) Damour-Sch\"afer test. In this work, we utilize the exquisite timing precision of PSR~J1713+0747 and the tight limit on $\dot{e}$ 
to directly test the violation of UFF in the strong-field regime.

Based on the equation of motion, \citet{ds91} derived that the eccentricity vector of the 
pulsar orbit would change according to (neglecting terms of order $e^2$ and smaller)
\begin{equation}\label{eq:edot}
    \dot{\mathbf e} \simeq \frac{3}{2\msV_O}\,\Delta\,
      {\mathbf g} \times\hat{\mathbf k} +
      \dot{\omega}_{\rm PN}\,\hat{\mathbf k} \times {\mathbf e} \,,
\end{equation}
for given violation parameter of UFF, $\Delta$. Here the vector ${\mathbf e}$ points at 
periastron, $\hat{\mathbf k}$ is a unit vector parallel to orbital angular momentum, 
$\msV_O \equiv [\msG (\Mp+\Mc) n_{\rm b}]^{1/3}$ is the relative 
orbital velocity between the two stars. The second term describes the post-Newtonian periastron advance rate:
\begin{equation}\label{eq:padv}
    \dot{\omega}_{\rm PN} \simeq 3\msF (\msV_O/c)^2 n_{\rm b} + {\cal O}(e^2) \,,
\end{equation}

\begin{figure}
    \includegraphics[width=\columnwidth]{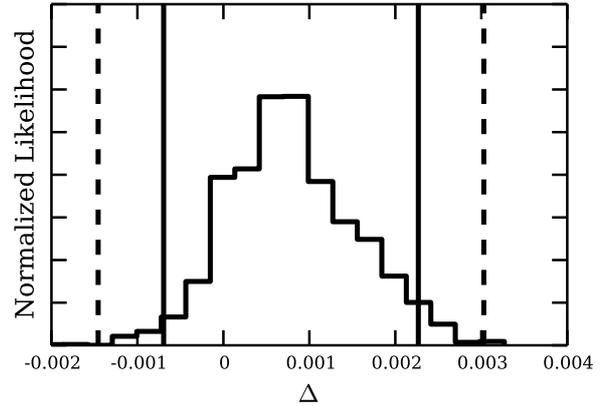} \\ 
    \caption {\label{fig:delta} The normalized likelihood distribution of $\Delta$ derived from the MCMC of PSR J1713+0747's timing and noise parameters. Solid line is the 95\% confidence limit, dashed line marks the 99\% confidence limit.}
\end{figure}

The extended {\it ELL1} timing model allows us to measure the change rate of the eccentricity vector, and 
we do detect an apparent $\dot{e}_y = (-1.7 \pm 0.4) \times 10^{-17}$~s$^{-1}$ along 
the line of sight, likely coming from the periastron advance of the orbit. After 
removing the contributions of periastron advance ($\dot{e}_x^{\rm PN} = -0.07 \times10^{-17}$~s$^{-1}$, $\dot{e}_y^{\rm PN} = -1 \times10^{-17}$~s$^{-1}$) according to the measured system masses and 
orbital parameter (Table~\ref{tab:pars}) and eq.~(\ref{eq:padv}), the resulting excess eccentricity change rate [$\dot{e}_x^{\rm exc}=(0.4\pm0.4)\times10^{-17}$~s$^{-1}$, $\dot{e}_y^{\rm exc}=(0.7\pm0.4)\times10^{-17}$~s$^{-1}$] is consistent with zero. We perform a Monte-Carlo Markov Chain (MCMC) simulation that simultaneously fits
both the timing model and the dispersion, jitter and red noises, using the PAL2 software. The MCMC allows us to obtain a 
large sample of possible timing parameter values along with their likelihood. We then use 
these MCMC results to calculate the $\Delta$ needed to account for the residual $\dot{e}$ 
excess. Figure~\ref{fig:delta} shows the posterior distribution of $\Delta$.
From this result we derive that $-0.0007 <\Delta<0.0023$ with 95\% confidence level (C.L.). The deviation from GR is insignificant. For the ease of comparison 
with previous results, we also derive from the above results:
\begin{equation}\label{eq:UFFlimit}
  |\Delta| < 0.002 \quad \mbox{(95\% C.L.)} \,.
\end{equation}
This constraint improves the previous best pulsar test of UFF in gravitation 
\citep{gsf+11} by more than a factor of two. More importantly, it is a direct test and 
therefore, as discussed above, more robust than previous Damour-Sch\"afer 
test based limits. We discuss the theoretical meaning of limit in eq. ~(\ref{eq:UFFlimit}) in Section~\ref{sec:summary}. 

\subsection{Testing the Lorentz Invariance and Conservation of Momentum in Gravitation}
\label{sec:alpha3}

The PPN parameters $\alpha_1$, $\alpha_2$, and $\alpha_3$ describe different symmetry 
breaking effects in gravitational theories that violate local Lorentz invariance in the
gravitational sector. The parameters $\alpha_1$ and $\alpha_2$ have already been tightly 
constrained through binary and isolated pulsars \citep{sw12, sck+13}. In this paper, 
we focus on the test of $\alpha_3$, but we also account for those effects related to 
$\alpha_1$ and $\alpha_2$ which in principle could have an influence on our $\alpha_3$ 
limits.

The parameter $\alpha_3$ describes a gravitational symmetry breaking that leads to 
both the existence of a preferred frame and a violation of conservation of momentum 
\citep{Will93}. A non-zero $\alpha_3$ would give rise to an anomalous self-acceleration for a 
spinning self-gravitating body that moves in the preferred reference frame. 
One often assumes that the universal matter distribution selects the rest frame of 
the Cosmic Microwave Background (CMB) as the preferred frame. 
Following the same idea, we choose CMB frame as the preferred frame in our analysis. However, our constraint on $\hat{\alpha}_3$ should be robust for preferred frames that moves at low velocity with respect to the pulsar, and it gets stronger when the selected frame is moving very fast with respect to the pulsar.
A weakly self-gravitating body with mass $M$, gravitational binding energy $E_{\rm grav}$, 
rotational frequency $\nu$, and velocity ${\bf v}_{\rm CMB}$ in
the CMB frame, would undergo acceleration induced by $\alpha_3$ effects: 
\begin{equation} \label{eq:a3}
  {\mathbf a}_{\alpha_3} = -\frac{\alpha_3}{3} \, \frac{E_{\rm grav}}{M c^2} \,
    2\pi\nu \, \hat{\mathbf n}_s \times {\bf v}_{\rm CMB} \,.
\end{equation}
Here $\hat{\bf n}_s$ is a unit vector in the direction of the body's spin \citep{Will93}. For 
strongly self-gravitating bodies, following \cite{bd96}, we replace 
$E_{\rm grav}/M c^2$ by the sensitivity of the pulsar $s_{\rm p}$.\footnote{\cite{bd96} 
introduced the compactness $c_{\rm p} = 2 s_{\rm p}$.} 
Furthermore, we replace $\alpha_3$ by $\hat\alpha_3$, where the hat symbol serves as a 
reminder that we are testing the (body-dependent) strong-field extension of $\alpha_3$, 
which may be different from the weak-field $\alpha_3$, but is expected to be of 
the same order (see Section~\ref{sec:summary} for a more detailed discussion). 

Similar to the violation of UFF, the acceleration caused by $\hat{\alpha}_3$ would lead to 
a polarization of the orbit of a rapidly rotating pulsar \citep{bd96}. For fully 
recycled millisecond pulsars like PSR~J1713+0747, $\hat{\bf n}_s$ likely 
aligns with the orbital angular momentum. For this reason, we can set 
$\hat{\bf n}_s = \hat{\bf k}$ in our calculations.

Besides the influence of a non-vanishing $\alpha_3$, one also needs to account for 
contributions of $\alpha_1$ and $\alpha_2$ to the orbital dynamics, since in 
general one cannot assume $\alpha_1$ and $\alpha_2$ to be zero in a theory that
breaks local Lorentz invariance in the gravitational sector and gives rise to a 
$\alpha_3$. In a near-circular binary, a non-vanishing $\alpha_2$ would lead 
to a precession of the orbital angular momentum around the direction of ${\bf w}$, which 
in turn leads to a temporal change in the projected semi-major axis $x$ \citep{sw12}. On 
the other hand $|\alpha_2|$ is constrained by pulsar experiments to be less than about 
$10^{-9}$, which corresponds to a change in $x$ that is about six orders of magnitude 
smaller than the contribution from proper motion, which is used to determine $\Omega$. 
Therefore effects from $\alpha_2$ can be safely neglected in our $\alpha_3$ test. 

A non-vanishing $\alpha_1$ adds to the polarization of the orbit in the same way as a 
non-vanishing $\alpha_3$, and analogous to eq.~(\ref{eq:edot}) one finds for the 
change of the orbital eccentricity vector
\begin{equation} \label{eq:edot_a1a3}
    \dot{\mathbf e} =  
      \dot{\mathbf e}_{\hat{\alpha}_1} +
      \dot{\mathbf e}_{\hat{\alpha}_3} +
      \dot{\omega}_{\rm PN}\,\hat{\mathbf k} \times {\mathbf e} \,,
\end{equation}
where 
\begin{equation} \label{eq:edot_a1}
    \dot{\mathbf e}_{\hat{\alpha}_1} \simeq \frac{\hat{\alpha}_1}{4c^2} \,
      \frac{\Mp - \Mc}{\Mp + \Mc} \, n_{\rm b} \msV_O \, {\bf w}_\perp
\end{equation}
\citep{de92a} and
\begin{equation} \label{eq:edot_a3}
  \dot{\mathbf e}_{\hat{\alpha}_3} \simeq
    \frac{3}{2\msV_O}\,{\mathbf a}_{\hat{\alpha}_3} \times\hat{\mathbf k} =
    -\hat{\alpha}_3 \pi \, \frac{s_{\rm p}\nu}{\msV_O} \, {\bf w}_\perp 
\end{equation}
\citep{bd96}. The velocity ${\bf w}_\perp$ is the projection of the systemic CMB 
frame velocity into the orbital plane of the binary pulsar. To independently 
constrain $\hat\alpha_3$ from PSR~J1713+0747 timing, we will include the $\hat{\alpha}_1$ 
limits obtained from PSR~J1738+0333 by \cite{sw12} in our analysis. 
Conversely, owning to its small orbital period, PSR~J1738+0333's possible $\hat\alpha_3$ effects would be much smaller than its $\hat\alpha_1$ effects and would have insignificant impact on the $\hat{\alpha}_1$ limits derived from this system. 
There is one assumption, however, that we have to make 
here. Since PSR~J1713+0747 and PSR~J1738+0333 have different masses, and therefore 
different sensitivities $s_{\rm p}$, we cannot assume that they would lead to identical 
$\hat{\alpha}_1$. For this reason, our analysis only applies to deviations from GR which 
exhibit only a moderate mass dependence of the strong-field parameter $\hat{\alpha}_1$, at 
least for neutron stars in the range of 1.3 to 1.5\,$\Msun$.

\begin{figure}
  \includegraphics[width=\columnwidth]{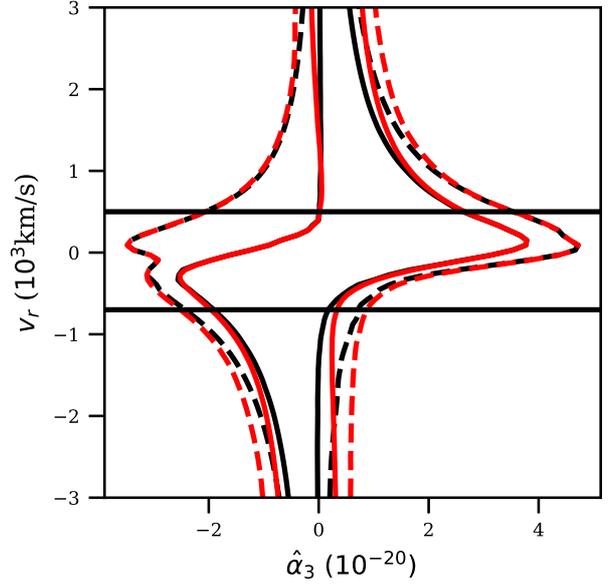} \\ 
  \caption {\label{fig:alpha3} The likelihood distribution of $\hat{\alpha}_3$ 
    derived from the MCMC of PSR J1713+0747's timing and noise parameters as a
    function of the assumed line-of-sight velocity $v_r$. The black solid curves 
    corresponds to the 95\% C.L. and the black dashed line to the 99\% C.L. The red curves shows the same limits with $\hat{\alpha}_1$ taken into 
    account (see Section \ref{sec:alpha3} for details). The two horizontal lines labels 
    the escape velocity of the pulsar binary.} 
\end{figure}

All parameters involved in the evaluation of $\dot{\mathbf e}$ are 
measurable through the timing of PSR~J1713+0747 (Table~\ref{tab:pars}), except for the radial velocity of the pulsar binary with respect to the Solar system $v_r$. To calculate $\dot{\bf e}$ from eq.~(\ref{eq:edot_a1a3}) one needs the 
system's three-dimensional (3D) velocity ${\mathbf v}_{\rm CMB}$ in the CMB rest frame. ${\mathbf v}_{\rm 
CMB}$ can in principle be computed from the binary's 3D velocity $\mathbf v$ about
our Solar System by adding the Solar System speed in the CMB rest frame, which is well 
known from the measurement of the CMB dipole 
\citep[see][for the latest measurement]{aaa+13}. We measure the pulsar's proper motion and distance 
(Table \ref{tab:pars}), which allows us to determine the transverse component of ${\bf v}$. 
The white dwarf companion of PSR\,J1713+0747 is relatively faint \citep{lfc96}, and a measurement of the radial velocity $v_r$ through optical spectroscopy is currently not available.
Therefore, we treat $v_r$ as a free parameter and calculate 
the limits on $\hat{\alpha}_3$ as a function of $v_r$. A limit for $v_r$ comes from the 
plausible assumption that the PSR~J1713+0747 system is bound to the Galaxy and therefore must 
be slower than the Galactic escape velocity. Taking the Galactic potential of 
\cite{McMillan16}, we find $v_r$ to be within the range of about $-680$ to 
$+460$\,km/s.\footnote{As a cross-check we also used the potential by \cite{kbgb08}, which 
gives very similar results.} 
As shown in Figure~\ref{fig:alpha3}, the constraint on $\hat{\alpha}_3$
tightens as $|v_r|$ gets larger because large $v_r$ lead to large ${\bf v}_{\rm CMB}$,
which enhance the polarization effects. We find
that in the most conservative scenario by taking the minimum $\hat{\alpha}_3$ value from the left-side of the 95\% C.L. contour as the lower bound and the maximum value from the right-side of the contour as the upper bound (despite the fact that the two values correspond to different values of $v_r$):
\begin{equation} \label{eq:a3limit}
  -3 \times 10^{-20} < \hat{\alpha}_3 < 4 \times 10^{-20} \quad
  \mbox{(95\% C.L.)} \,.
\end{equation}
This result is better than the previous best constraint on 
$\hat{\alpha}_3$, $|\hat{\alpha}_3|<5.5\times 10^{-20}$ from \citep{gsf+11}, which is based on a Damour-Sch\"afer type of test using an ensemble of pulsars. 

\section{Discussion}
\label{sec:summary}

In this paper, we present new tests of the constancy of $G$, and the violations of UFF, 
Lorentz invariance, and conservation of momentum in gravitation. These violations of 
gravitational symmetries lead to changes in the orbital period and eccentricity particularly in binaries such as pulsar-white dwarf systems like J1713+0747. We conduct 
these tests through the measurement of an excess change in the orbital period and 
eccentricity from the timing analysis of PSR~J1713+0747.

We repeat the combined $\dot{G}/G$, $\kappa_{\rm D}$ test presented in \citet{zsd+15} by 
incorporating more PSR~J1713+0747 timing data from EPTA and find the result improves
[$\dot{G}/G = (-0.1\pm0.9)\times10^{-12}$~yr$^{-1}$ and $\kappa_{\rm D} = 
(-0.7\pm2.2)\times10^{-4}$]. The improvement on $\dot{G}/G$ could be attributed to the 
inclusion of EPTA data which increased the number of TOAs used in the experiment, whereas the changes in $\kappa_{\rm D}$ limits 
are mostly due to the change from using a fiducial sensitivity function that scales 
linearly with neutron star mass to using a more realistic non-linear neutron star 
sensitivity function. Our $\kappa_{\rm D}$ test is a generic test of dipolar gravitational 
radiation, included for the purpose of generalizing the $\dot{G}/G$ test for general 
SEP-violating theories. More stringent tests of the dipolar gravitational radiation effects 
could be done with pulsar timing if one takes into account the nature of the SEP-violating theories \citep{fkw12}. In some theories, ${\cal O}(s_{\rm p}^3)$ terms cannot be 
neglected (cf.\ eq.~(\ref{eq:PbdotD})) even in a first order estimation.  
In some theory-specific tests, a more stringent limit on $\kappa_{\rm D}$ 
could come from non-radiative tests, for instance from the Solar system measurements based on the Cassini experiment, like 
for JFBD gravity where Cassini implies $|\kappa_{\rm D}| \lesssim 9 \times 10^{-5}$ with 
95\% confidence (cf.\ eq.~(\ref{eq:kappaD_JFBD})).

When compared directly to the Solar System tests (LLR \citep{hmb10} and planetary 
ephemerides \citep{fle+15}; Figure \ref{fig:gdot}), our $\dot{G}/G$ constraint is not as 
tight. But the pulsar binary tests involve objects of much stronger self-gravitation 
than objects in the Solar System. The pulsar timing limits of $\dot{G}/G$ and 
$\kappa_{\rm D}$ are testing SEP-violating effects beyond linear extrapolations from the 
weak-field limit. \citet{wex14} demonstrated that in certain theories of gravitation, $\dot{G}/G$  effects could be greatly enhanced by a strongly self-gravitating body while remaining insignificant in the Solar System. 

Previous pulsar UFF tests, such as \cite{wex97,sfl+05} and \cite{gsf+11} employed the idea of \citet{ds91}, which uses an ensemble of wide-orbit 
small-eccentricity pulsar-white dwarf binaries. The effectiveness of that approach relies on the smallness of $e$ and the statistical argument that the unknown orientations of the orbits from a collection of 
pulsars are randomly and uniformly distributed. Hence the \citet{ds91} tests cannot 
directly detect SEP violation and may only improve when a new pulsar binary with a better figure of merit is found. There comes a further caveat with previous tests, as these tests are based on an ensemble of systems with different neutron-star masses. Hence, for this mix of neutron-star masses a priori assumptions about the strong-field behavior of gravity had to be made \citep{dam09}. For our tests this is not the case, since these tests are based on a single neutron-star with well determined mass.

\citet{ds91} also pointed out the possibility of directly testing UFF violations 
by constraining temporal changes in the orbital eccentricity. Such a test has the 
advantage of not depending on a group of pulsar binaries, the smallness of their $e$ and 
assumptions on their orbital orientations. The effectiveness of this test improves with 
timing precision. \citet{fkw12} identified PSR~J1713+0747 as one of the best candidates 
for such a direct test. But at that time $\dot{\bf e}$ was not directly modeled in the 
timing of that pulsar, and \citet{fkw12} used an estimate of the upper limit of
$\dot{e}$ based on the uncertainties of the measured $e$. As a result, \citet{fkw12} could 
put some preliminary limits on UFF violations. In this paper, we 
conduct the first direct UFF test with a measured $\dot{\bf e}$, which in principle could detect a violation of UFF, should the effect be strong enough.  

Using an extended version of the {\it ELL1} timing model of \cite{lcw+01} in our 
analysis, we find $\dot{\bf e} \simeq (0.4 \pm 0.4,-1.7 \pm 0.4) \times 10^{-17}$\,s$^{-1}$ 
(see Table~\ref{tab:pars}), which is consistent with being caused by post-Newtonian 
periastron advance predicted by GR. We find no evidence for a violation of UFF with 
$|\Delta|<0.002$, a result that improves by more than a factor of 2 from the previous best 
constraint \citep{sfl+05,gsf+11}. Similarly, we find $-3\times10^{-20} < \hat{\alpha}_3 < 4\times10^{-20}$, which is also better than the previous best result. These limits go beyond the PPN framework since they also capture strong field deviations. To illustrate this, for example for 
$\alpha_3$, we expand $\hat\alpha_3$ with respect to the sensitivity
\begin{equation}\label{eq:a3exp}
  \hat\alpha_3 = \alpha_3 + \alpha_3^{(1)} s_{\rm p}  +
  \alpha_3^{(2)}  s_{\rm p}^2 + \dots\,,
\end{equation}
then the limit in eq.~(\ref{eq:a3limit}) not only constrains the weak-field
counterpart, $\alpha_3$, at the level of ${\cal O}(10^{-20})$, but also
poses strong constraints on higher-order terms, $\alpha_3^{(1)}$,
$\alpha_3^{(2)}$, and so on. The same applies for the strong-field generalization of the 
Nordtvedt parameter, that is given by $\Delta \equiv \hat\eta_{\rm N} s_{\rm p}$. Detailed 
accurate mapping of strong-field generalization and weak-field counterpart needs explicit 
calculations in specified gravity theories. Note, equations like (\ref{eq:a3exp}) 
ultimately fail to capture non-perturbative strong field deviations, like spontaneous 
scalarization \citep{de93}.

Our UFF test using PSR~J1713+0747 is expected to be surpassed by tests using the
recently discovered pulsar in a stellar triple system \citep{rsa+14}, as that system has a 
much stronger external gravitational field provided by the third star. Simulations 
suggest an improvement by at least three orders of magnitude \citep{bbc+15,shao16,kra16}. 
However, PSR~J1713+0747 will remain to be one of the best systems for testing $\dot{G}$
with pulsars, due to its wide orbit and high timing precision, and for testing 
$\hat{\alpha}_3$ because of the fast rotation of this pulsar and its well-constrained 
orbit. 

\section*{Acknowledgements}
WWZ is supported by  the Chinese Academy of Science Pioneer Hundred Talents Program and the Strategic Priority Research Program of the Chinese Academy of Sciences\ Grant No. XDB23000000.
Pulsar research at UBC is supported by an NSERC Discovery Grant and
Discovery Accelerator Supplement, and by the Canadian
Institute for Advanced Research. GD, MK and KL acknowledge financial support by the European Research Council (ERC) for the ERC Synergy Grant BlackHoleCam under contract no. 610058. 
GJ and GS acknowledge support from the Netherlands Organisation for Scientific Research NWO (TOP2.614.001.602).
JAE acknowledges support
by National Aeronautics and Space Administration (NASA) through Einstein Fellowship grant PF4-150120. Some computational work was performed on the Nemo cluster at UWM supported by National Science Foundation (NSF) grant No. 0923409. 
CGB acknowledges support from the European Research Council under the European Union's Seventh Framework Programme (FP/2007-2013) / ERC Grant Agreement nr. 337062 (DRAGNET).
KJL is supported by the strategic Priority Research Program of Chinese Academy of Sciences, Grant No. XDB23010200,NSFC 2015CB857101, U15311243.
SO acknowledges support from the Alexander von Humboldt Foundation and through the ARC Laureate Fellowship grant FL150100148.
Portions of this research were carried out at the Jet Propulsion Laboratory,
California Institute of Technology, under a contract with NASA. The
NANOGrav work in this paper was supported by National Science Foundation (NSF) PIRE program award number 0968296 and NSF Physics Frontiers Center award number 1430824.
This work was supported by NSF grant 0647820. The Arecibo
Observatory is operated by SRI International under a cooperative
agreement with the NSF (AST-
1100968), and in alliance with Ana G. Méndez-Universidad
Metropolitana, and the Universities Space Research Association.
The Green Bank Observatory is a facility of the National Science Foundation operated under cooperative agreement by Associated Universities, Inc.
Part of this work is based on observations with the 100-m telescope of the
Max-Planck-Institut f\"ur Radioastronomie (MPIfR) at Effelsberg in Germany. Pulsar
research at the Jodrell Bank Centre for Astrophysics and the observations using
the Lovell Telescope are supported by a consolidated grant from the STFC in the
UK. The Nan{\c c}ay radio observatory is operated by the Paris Observatory,
associated to the French Centre National de la Recherche Scientifique (CNRS), and to the Universit\'e d'Orl\'eans. The Westerbork Synthesis Radio Telescope
is operated by the Netherlands Institute for Radio Astronomy (ASTRON) with
support from the Netherlands Foundation for Scientific Research (NWO). 
The authors acknowledge the use of the Hercules computing cluster from Max-Planck Computing and Data Facility.


\bibliographystyle{mnras}
\bibliography{journals,myrefs,NWrefs,modrefs,psrrefs,crossrefs}

\appendix

\section{Galactic corrections}
\label{sec:appA}

In Section \ref{sec:gdot}, we discuss how the changing Doppler effect related to the 
systemic motion of a binary pulsar causes apparent variations in the orbital parameters. 
An important part of the changing Doppler effect comes from the Galactic contribution to the 
line-of-sight acceleration between the pulsar binary and the Solar System. There are two orthogonal components in this Galactic contribution: the difference in the centrifugal acceleration in our circular motion around the Galactic center, and the difference in the vertical acceleration caused mainly by the Galactic disk. We can derive the expected $\dot{P}^{\rm Gal}_{\rm b}/P_{\rm b}$ based on our knowledge of the Galactic rotation curve and the local surface density of the Galactic disk.   

\begin{figure}
    \includegraphics[width=\columnwidth]{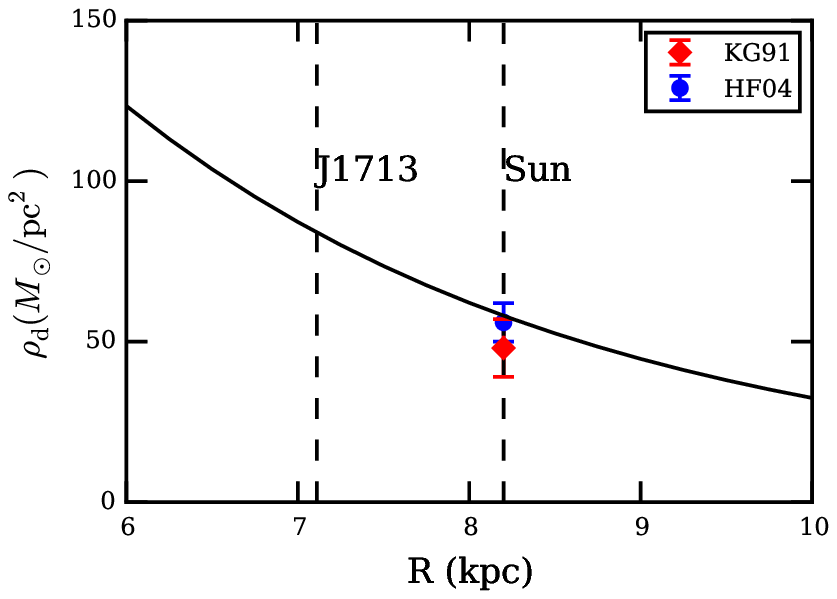} \\ 
    \caption {\label{fig:GalDen} The solid curves show the models of Galactic disk surface density  as a function of Galactic radius from \citet{McMillan16}. The error bars indicate disk surface densities from studies of the dynamics of stars in the Solar vicinity by \citet{kg91} (red diamond) and \citet{hf04a} (blue circle). The vertical lines mark the positions of the pulsar and the Sun.}
\end{figure}

Based on \citet{dt91}, \citet{nt95} derived an analytical formula for calculating the apparent orbital variation using a flat rotation curve and the vertical acceleration model for the Solar vicinity. The subsequent pulsar timing works \citep{dvtb08,vbv+08,lwj+09,fwe+12,zsd+15} used the same formula with the most recent Galactic disk model coming from \citealt{hf04a} and \citealt{rmb+14}. However, the analytical approach put forward by \citet{dt91,nt95} is a good approximation only when the pulsar systems are close to the Solar System or have a comparable distance to the Galactic center. As shown in Figure \ref{fig:GalDen}, J1713+0747 is about 1~kpc closer to the Galactic center than the Sun. Therefore, it experiences a slightly higher vertical gravity than modeled previously in \citet{zsd+15}. 

In this work, we employ a new Galactic model by \citet{McMillan16}, which fits the rotation curve and the stellar dynamic data with an axisymmetric Galactic potential and provides a code for computing the Galactic gravitational acceleration. It is worth noting that the key input data for constraining their model regarding vertical forces were from \citealt{kg91} --- an earlier study than \citealt{hf04a}. But as shown in Figure \ref{fig:GalDen}, the later \citealt{hf04a} result fits \citet{McMillan16} model better and the changes were relatively small. Table \ref{tab:pbdot} shows J1713+0747's $\dot{P}^{\rm Gal}_{\rm b}$ computed using various Galactic potential models and the observed excess after removing Shklovskii and GR effects. One can see that most realistic models \citep{db98,bt08,Piffl14,McMillan16} fit the observed excess better than the analytical approximation.

\begin{figure}
    \includegraphics[width=\columnwidth]{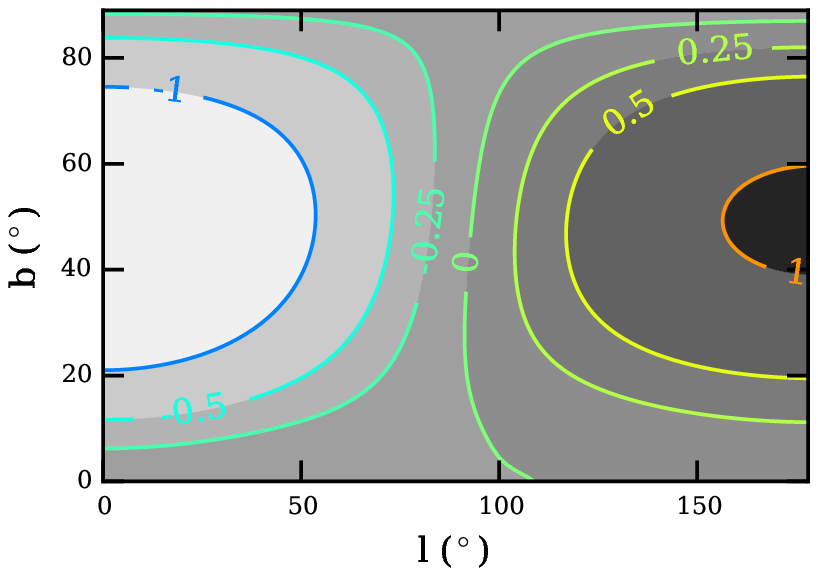} \\ 
    \caption {\label{fig:AccDiff1} The difference between the Galactic corrections ($\dot{P}^{\rm Gal}/P$ and $\dot{P}^{\rm Gal}_{\rm b}/P_{\rm b}$ ) derived from \citet{McMillan16} and the \citet{nt95} approximation as a function of Galactic longitude ($l$) and latitude ($b$) for a pulsar binary at 1\,kpc distance from us in units of $10^{-12}$\,yr$^{-1}$.}
\end{figure}

Figure \ref{fig:AccDiff1} shows the $\Delta\dot{P}^{\rm Gal}_{\rm b}/P_{\rm b}$ from using the \citealt{McMillan16} model instead of \citealt{nt95} for a putative pulsar binary at 1~kpc distance as a function of Galactic longitude and latitude. 
The correction becomes of the same order of magnitudes as the observed effect $|\dot{P}_{\rm b}/P_{\rm b}|$ ($\sim10^{-12}$~yr$^{-1}$) for J1713+0747. PSR J0437-4715 is another pulsar binary system that is sensitive to this effect \citep{vbv+08, dvtb08,rhc+16}. However, at a distance of only 0.16~kpc, this pulsar's Galactic acceleration could be computed fairly accurately by \citealt{nt95} model, with a $|\Delta\dot{P}^{\rm Gal}_{\rm b}/P_{\rm b}|<0.2\times10^{-12}$~yr$^{-1}$. 
For the other pulsar binary used in our $\dot{G}/G$ analysis --- PSR J1738+0333, the correction is similar to that of J1713+0747, but still relatively insignificant compared to the observational uncertainty $\delta\dot{P}_{\rm b}/P_{\rm b}$ of $\sim4\times10^{-12}$ \citep{fwe+12}. 

\begin{figure}
    \includegraphics[width=\columnwidth]{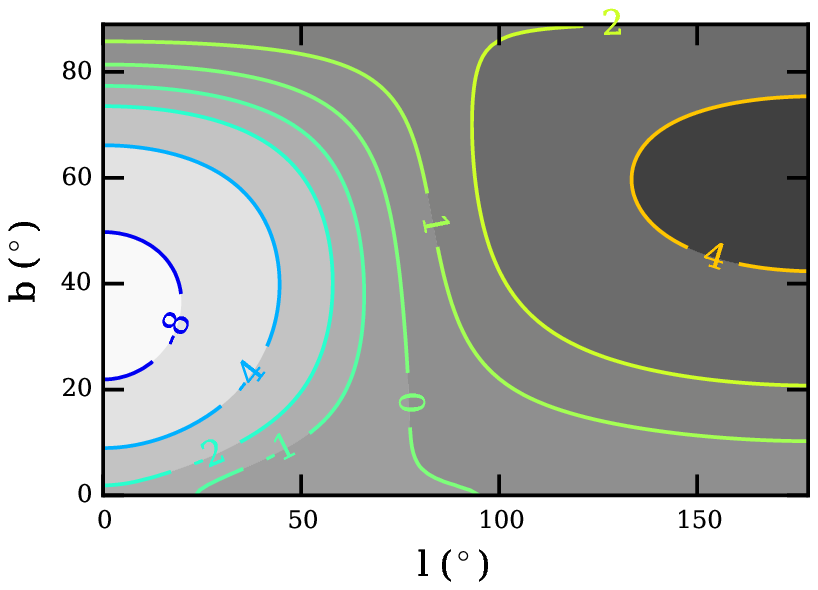} \\ 
    \caption {\label{fig:AccDiff3} The difference between the Galactic corrections ($\dot{P}^{\rm Gal}/P$ and $\dot{P}^{\rm Gal}_{\rm b}/P_{\rm b}$ ) derived from \citet{McMillan16} and the \citet{nt95} approximation as a function of Galactic longitude ($l$) and latitude ($b$) for a pulsar binary at 3\,kpc distance in units of $10^{-12}$\,yr$^{-1}$.}
\end{figure}

Since \citealt{nt95} used the local disk surface density and assumed a flat rotation curve, the real Galactic potential deviates more from it at greater distances from the Sun.
In Figure \ref{fig:AccDiff3}, we show that at a distance of 3~kpc the extra Galactic correction $|\Delta\dot{P}_{\rm b}/P_{\rm b}|$ (or $|\Delta\dot{P}/P|$ for the pulsar period) comes close to $10^{-11}$~yr$^{-1}$ for some directions, therefore, a more realistic model must be used for pulsars in this situation.

Presently, there are only a handful of pulsars that are sensitive to the Galactic acceleration. In the future, new telescopes such as the Five hundred meter Aperture Spherical Telescope (FAST; \citealt{FAST,lp16}) and the Square Kilometer Array (SKA; \citealt{ks15}) could improve both the timing and distance measurement for many more pulsars such that this correction becomes significant for them. It might be possible to start using some pulsars as accelerometers for probing the Galactic gravity field and improving our knowledge of the Galactic potentials, independent to the improvements expected from GAIA\footnote{http://sci.esa.int/gaia/} \citep{Gaia} . 
  
\begin{table}
\caption{PSR J1713+0747's $\dot{P}^{\rm Gal}_{\rm b}$ components in units of 
$10^{-12}\,{\rm s\,s}^{-1}$,
in comparison to the observed $\dot{P}_{\rm b}$ where only the Shklovskii contribution has
been subtracted.}
\label{tab:pbdot}
\begin{tabular}{lccc}
\hline
Galactic potential/model & horizontal & vertical & total  \\
\hline
 \citet{McMillan16} & 0.16 & $-$0.50 &  $-$0.34 \\
 \citet{Piffl14}    & 0.14 & $-$0.46 &  $-$0.33 \\
 \citet{bt08}       & 0.16 & $-$0.43 &  $-$0.27 \\
 \citet{db98}       & 0.17 & $-$0.46 &  $-$0.29 \\
 \citet{nt95}$^*$   & 0.27 & $-$0.36 &  $-$0.10 \\
\hline
$\dot{P}^{\rm obs}_{\rm b} - \dot{P}^{\rm Shk}_{\rm b}$& --- & --- & $-$0.31(15)\\
\hline
\end{tabular}
\begin{tabular}{l}
$^*$ Analytical model including updates from \\ \citet{lwj+09,fwe+12,zsd+15}.
\end{tabular}
\end{table}

\label{lastpage}
\end{document}